\documentclass[aps,preprint,nofootinbib,showpacs]{revtex4}
\usepackage[dvips]{graphicx,color}
\usepackage{amsmath}
\usepackage{cases}
\usepackage{color,ulem}

\newcommand{\be}{\begin{equation}}
\newcommand{\ee}{\end{equation}}
\newcommand{\ba}{\begin{eqnarray}}
\newcommand{\beq}{\begin{equation}}
\newcommand{\eeq}{\end{equation}}
\newcommand{\ea}{\end{eqnarray}}

\newcommand{\MNS}{{\text{MNS}}}

\newcommand{\eV}{\text{eV}}
\newcommand{\MeV}{\text{MeV}}
\newcommand{\GeV}{\text{GeV}}
\newcommand{\TeV}{\text{TeV}}
\newcommand{\fb}{\text{fb}}

\newcommand{\BR}{\text{BR}}

\newcommand{\mLll}{{(m_L^{})_{\ell\ell^\prime}}}
\newcommand{\mDil}{{(m_D^{})_{i\ell}}}
\newcommand{\hll}{{h_{\ell\ell^\prime}^{}}}
\newcommand{\qqHppHmm}{{q\overline{q}\to \gamma^*,Z^* \to H^{++}H^{--}}}

\newcommand{\qqHpmpmHmp}{{q^\prime\overline{q}\to W^* \to H^{\pm\pm}H^\mp}}

\def\beqa{\begin{eqnarray}}
\def\eeqa{\end{eqnarray}}
\newcommand{\SU}{{\text{SU}}}
\newcommand{\U}{{\text{U}}}

\def\bea{\begin{eqnarray}}
\def\eea{\end{eqnarray}}

\def\err#1#2{\lower2pt\hbox{ $\stackrel{\scriptstyle +#1}{\scriptstyle -#2}$}}
\def\ga{\mathrel{\raise.3ex\hbox{$>$\kern-.75em\lower1ex\hbox{$\sim$}}}}
\def\la{\mathrel{\raise.3ex\hbox{$<$\kern-.75em\lower1ex\hbox{$\sim$}}}}
\def\bmaT{\left(\begin{array}{ccc}}
\def\emaT{\end{array}\right)}
\def\bma{\left( \begin{array} }
\def\ema{\end{array} \right)}
\def\gsim{~{\rlap{\lower 3.5pt\hbox{$\mathchar\sim$}}\raise 1pt\hbox{$>$}}\,}
\def\lsim{~{\rlap{\lower 3.5pt\hbox{$\mathchar\sim$}}\raise 1pt\hbox{$<$}}\,}

\begin{document}

\preprint{
\vbox{%
\hbox{MISC-2013-06}
}}
\title{\boldmath
 Dependence of the leptonic decays of $H^\pm$
on the neutrino mixing angles $\theta_{13}$ and $\theta_{23}$
in models with neutrinophilic charged scalars
\unboldmath} 
\author{A.G.~Akeroyd}
\email{a.g.akeroyd@soton.ac.uk}
\author{S.~Moretti}
\email{S.Moretti@soton.ac.uk}
%\altaffiliation{}
\affiliation{School of Physics and Astronomy, University of Southampton \\
Highfield, Southampton SO17 1BJ, United Kingdom}
\affiliation{Particle Physics Department, Rutherford Appleton Laboratory, 
Chilton, Didcot, Oxon OX11 0QX, United Kingdom}
\author{Hiroaki Sugiyama}
\email{sugiyama@cc.kyoto-su.ac.jp}
\affiliation{Maskawa Institute for Science and Culture,
Kyoto Sangyo University, Kyoto 603-8555, Japan}

\begin{abstract}
In the Higgs Triplet Model and the neutrinophilic Two-Higgs-Doublet Model the 
observed neutrinos obtain mass from a vacuum expectation value
which is much smaller than the vacuum expectation value
of the Higgs boson in the Standard Model. Both models contain a singly charged Higgs boson~($H^\pm$)
whose Yukawa coupling is directly related to the neutrino mass (i.e.\ a "neutrinophilic charged Higgs").
 The partial decay widths of $H^\pm$
into a charged lepton and a neutrino~($H^\pm\to \ell^\pm\nu$)
depend identically on the neutrino masses and mixings in the two models.
 We quantify the impact of the recent measurement of $\sin^2{2\theta_{13}}$, 
which plays a crucial role in determining the
magnitude of the branching ratio of $H^\pm \to e^\pm\nu$
for the case of a normal neutrino mass ordering
if the lightest neutrino mass $m_0< 10^{-3}\,\eV$.
 We also discuss the sizeable dependence of $H^\pm \to \mu^\pm\nu$ and $H^\pm \to \tau^\pm\nu$ on $\sin^2{\theta_{23}}$, which would enable information to be obtained on
$\sin^2{\theta_{23}}$ and the sign of $\Delta m^2_{31}$ if these decays are measured.
Such information would help neutrino oscillation experiments
to determine the CP-violating phase $\delta$.

\end{abstract}
%14.80.Ec Other neutral Higgs bosons 
%14.80.Fd Other charged Higgs bosons
%12.60.Fr Extensions of electroweak Higgs sector
%14.60.Pq Neutrino mass and mixing
%\keywords: Higgs boson, Neutrino mass and mixing
\maketitle
%%%%%%%%%%%%%%%%%%%%%%%%%%%%%%%%%%%%%%%%%%%%%%%%%%

%%%%%%%%%%%%%%%%%%%%%%%%%%%%%%%%%%%%%%%%%%%%%%%%%%%%%%%%%%%%%%%%%%%%%%

%%%%%%%%%%%%%%%%%%%%%%%%%
%%%%  sec: intro   %%%%%%
%%%%%%%%%%%%%%%%%%%%%%%%%
\section{Introduction} 
\noindent
 The ATLAS~\cite{Aad:2012tfa} and CMS~\cite{Chatrchyan:2012ufa} experiments
at the CERN Large Hadron Collider~(LHC)
have discovered a new boson with a mass of approximately $125\,\GeV$. 
 The measurements of its branching ratios~(BRs) are consistent
(within experimental error)
with those predicted by the Higgs boson~\cite{Ref:Higgs}
of the Standard Model~(SM).
 Current LHC data~\cite{Chatrchyan:2012jja}
also suggests that the new particle's spin and parity
are compatible with the values expected for the SM Higgs boson.
 It is now widely believed that
this discovery corresponds to a fundamental scalar particle
with a vacuum expectation value~(vev)
i.e.\ it is a species of Higgs boson.
 Consequently,
there is increased motivation to search for additional scalars
which would belong to an extension of the SM
with a non-minimal Higgs sector.
 Such models might also provide a mechanism
for the generation of neutrino mass.
 Although the solitary Higgs boson in the SM
can provide a Dirac mass term  for the observed neutrinos
by assuming the existence of three generations of right-handed neutrinos,
such a mechanism would not be testable at the LHC\@.
Extensions of the Higgs sector of the SM
may involve an additional $\SU(2)_L$-multiplet of scalar fields
whose vev solely provides neutrino masses.
 We refer to these scalar fields as "neutrinophilic scalars".
 In this paper
we will consider two such models which are potentially testable
because they predict neutrinophilic charged scalars~($H^\pm$)
which might be light enough to be discovered at the LHC\@.

Neutrinos may obtain a Majorana mass
via the vev of a neutral Higgs boson
in an isospin triplet representation%
~\cite{Konetschny:1977bn, Mohapatra:1979ia,
Magg:1980ut,Schechter:1980gr,Cheng:1980qt}.
 A particularly simple implementation
of this mechanism of neutrino mass generation
is the "Higgs Triplet Model''~(HTM)
in which the SM Lagrangian is augmented
solely by an $\SU(2)_L$-triplet of scalar particles (denoted by $\Delta$)
with hypercharge $Y=2$%
~\cite{Konetschny:1977bn, Schechter:1980gr,Cheng:1980qt}.
 In the HTM
there are three electrically neutral Higgs scalars:
$h^0$ and $H^0$ are CP-even, and $A^0$ is CP-odd.
 These scalar eigenstates are
mixtures of the doublet and triplet neutral fields,
but the mixing angle is very small
in most of the parameter space of the HTM
because of the hierarchy of the vevs, $v_\Delta \ll v$,
where $v (=246\,\GeV)$ is the vev of the neutral doublet field, 
and $v_\Delta$ is the vev of the triplet field.
 There are also electrically charged scalars:
a doubly charged scalar~($H^{\pm\pm}$)
and a singly charged scalar~($H^{\pm}$).

 The Higgs sector of the SM may be extended
with a second $\SU(2)_L$-doublet scalar field of hypercharge $Y=1$
(denoted by $\Phi_\nu$)
which has a Yukawa interaction only with right-handed neutrinos.
The  phenomenology is discussed in Ref.~\cite{Ref:nuTHDM-M}
for the case where the right-handed neutrinos also have their Majorana mass terms%
~\cite{Ma:2000cc}.
 If right-handed neutrinos do not have Majorana masses%
~\cite{Ref:nuTHDM-D, Davidson:2009ha, Davidson:2010sf} then the
neutrinos are Dirac fermions,
and their mass matrix $(m_D)_{i\ell}^{}$ is solely given
by a product of new Yukawa coupling matrix $(y_\nu^{})_{i\ell}^{}$
and the vev $v_\nu^{}$ of the second scalar doublet.
 The vev $v_\nu^{}$ is generated
via spontaneous breaking of a global symmetry in Ref.~\cite{Ref:nuTHDM-D}
while it is obtained
via soft-breaking of a global symmetry
in Refs.~\cite{Davidson:2009ha, Davidson:2010sf}.
 We refer to the model of Dirac neutrinos
in Refs.~\cite{Davidson:2009ha, Davidson:2010sf}
as the "neutrinophilic Two Higgs Doublet Model"~($\nu$2HDM).
 Like the HTM,
the $\nu$2HDM also predicts
three electrically neutral Higgs scalars
(two being CP-even, and one being CP-odd),
as well as a singly charged scalar.

 In the context of both the HTM and the $\nu$2HDM
the simplest candidate for the observed boson at $\sim 125\,\GeV$
would be the lightest CP-even $h^0$.
 This scalar eigenstate has BRs which are
very similar to those of the SM Higgs boson
in most of the parameter space of the two models
with $v_\Delta^{}, v_\nu^{} \ll v$.
 At present,
the measured BRs of the $125\,\GeV$ boson are
fully consistent with those of the Higgs boson of the SM\@.
 The current experimental errors allow
deviations from the BRs of the SM Higgs boson
of the order of $20\%$ to $30\%$.
 The decay channel to two photons is sensitive to
the virtual effects of $H^\pm$ and $H^{\pm\pm}$%
~\cite{Arhrib:2011vc,Melfo:2011nx,Kanemura:2012rs,
Akeroyd:2012ms,Chun:2012jw,Dev:2013ff},
and the measurement of this decay now constrains
the parameters of the scalar potentials in the above models,
especially the mass of $H^{\pm\pm}$
and the trilinear coupling $h^0H^{++}H^{--}$ in the HTM\@.
The result of the ATLAS experiment~\cite{ATLAS_yy}
with all the data taken at $\sqrt s=7\,\TeV$ and $\sqrt s=8\,\TeV$ is
$R_{\gamma\gamma}=1.65\pm 0.24\text{(stat)}^{+0.25}_{-0.18}\text{(syst)}$,
where $R_{\gamma\gamma}=1$ for the SM Higgs boson.
 The CMS experiment measures
$R_{\gamma\gamma}=0.78\pm 0.27$ with a Multi-Variate-Analysis
and $R_{\gamma\gamma}=1.11\pm 0.31$ with a cut-based analysis~\cite{CMS_yy}.
 If future measurements show
a statistically significant deviation from $R_{\gamma\gamma} = 1$,
then this result could be readily explained by the presence of charged scalars. 

 The HTM and the $\nu$2HDM provide identical dependences of the partial decay widths for $H^{\pm\pm}\to \ell^\pm\nu$ on
the six neutrino oscillation parameters
and the unknown mass of the lightest neutrino,
where the main uncertainty comes from the latter parameter.
 Quantitative studies were performed
in the context of the HTM in Ref.~\cite{Perez:2008ha},
and in the $\nu$2HDM in Refs.~\cite{Davidson:2009ha,Davidson:2010sf}.
 In the HTM (in which the neutrinos are Majorana particles)
the prediction for
$\BR(H^\pm \to \ell^\pm \nu)$
is of particular importance
because its value does not depend on the two unknown Majorana phases
in the neutrino mass matrix.
 This result is in contrast to the prediction
for BR($H^{\pm\pm}\to \ell^\pm \ell^\pm$) in the HTM,
which {\it does} depend on the values of the Majorana phases
and thus such BRs have more uncertainty.
 Consequently, if a $H^{\pm\pm}$ and $H^\pm$ were discovered at the LHC,
a measurement of BR($H^{\pm}\to \ell^\pm\nu$) would provide
a more robust means of determining 
whether the mass of the neutrinos arose
solely from a triplet vev $v_\Delta$ (which is the case in the HTM),
or from a combination of mechanisms
which may or may not include a triplet vev.

In this work we study the dependence of $\BR(H^\pm \to \ell^\pm \nu)$
on the neutrino oscillation parameters,
in particular the mixing angles
$\sin^2{2\theta_{13}}$ and $\sin^2{\theta_{23}}$ of $U_\MNS$.
 Previous studies~\cite{Davidson:2009ha,Davidson:2010sf,Perez:2008ha}
considered the dependence of
$\BR(H^\pm \to \ell^\pm \nu)$ on these parameters by scanning over their allowed ranges
and presenting the results as scatter plots.
 The aim of the present work is to clarify the effect of
varying each of these mixing angles individually,
with special attention given to the impact of
the recent measurement of $\sin^2{2\theta_{13}}$.
We also pay attention to the dependence on $\sin^2{\theta_{23}}$,
whose uncertainty (whether $\sin^2{\theta_{23}} > 0.5$ or $\sin^2{\theta_{23}} < 0.5$)
might be the main hindrance in the determination
of the CP-violating phase $\delta$ in neutrino oscillation experiments.

Our work is organised as follows.
In section~II we briefly introduce the HTM and the 
$\nu$2HDM, and discuss the ongoing measurements of the neutrino
oscillation parameters. In section~III we present our numerical results for $\BR(H^\pm\to \ell^\pm\nu)$. Conclusions are given in section~IV.

%%%%%%%%%%%%%%%%%%%%%%%%%%%%%%%%%%%
%%%%%  sec: HTM and nu2HDMT %%%%%%%
%%%%%%%%%%%%%%%%%%%%%%%%%%%%%%%%%%%
\section{The Higgs Triplet Model and Neutrinophilic 2HDM}
The HTM and the $\nu$2HDM are models with a non-minimal Higgs sector in which 
the observed neutrinos obtain mass as a product of a Yukawa coupling and the vev
of a new scalar field. The two models predict the same specific relationship
between the neutrino parameters and the
partial widths of the decay channels $H^\pm\to \ell^\pm\nu$. In this section we briefly introduce both models, 
and then summarise the current experimental status of the
measurements of the neutrino oscillation parameters.

%%%%%%%%%%%%%%%%%%%%%
%%%  subsec: HTM  %%%
%%%%%%%%%%%%%%%%%%%%%
\subsection{HTM}
In the HTM~\cite{Konetschny:1977bn,Schechter:1980gr,Cheng:1980qt}
a $Y=2$ complex $\SU(2)_L$ isospin triplet of scalar fields,
${\bf T}=( T_1, T_2, T_3 )$, is added to the SM Lagrangian. 
This model has the virtue of providing Majorana masses for the observed neutrinos 
without the introduction of $\SU(2)_L$ singlet neutrinos.
The following $\SU(2)_L\otimes \U(1)$ gauge-invariant Yukawa interaction is introduced:
%-------------------------------
\begin{equation}
{\mathcal L}^{\text{HTM}}_{\text{Yuk}}
= -\hll L_\ell^TCi\sigma_2\Delta L_{\ell^\prime}+\text{h.c.}
\label{Eq:trip_yuk}
\end{equation}
%-------------------------------
 Here $\hll (\ell,\ell^\prime=e,\mu,\tau)$ is a complex
and symmetric coupling,
$C$ is the Dirac charge conjugation operator,
$\sigma_i (i=1\text{-}3)$ are the Pauli matrices,
$L_\ell=(\nu_{\ell L}, \ell_L)^T$ is a left-handed lepton doublet,
and $\Delta$  is a $2\times 2$ representation
of the $Y=2$ complex triplet fields:
%-------------------------------
\begin{equation}
\Delta
= {\bf T}\cdot\frac{\sigma}{2}
= T_1 \frac{\sigma_1}{2} + T_2 \frac{\sigma_2}{2} + T_3 \frac{\sigma_3}{2}
=\bma{cc}
\Delta^+/\sqrt{2}  & \Delta^{++} \\
\Delta^0       & -\Delta^+/\sqrt{2}
\ema ,
\label{Eq:Delta}
\end{equation}
%-------------------------------
where
$T_1 = (\Delta^{++} + \Delta^0)$,
$T_2 = i(\Delta^{++} - \Delta^0)$,
and $T_3 = \sqrt{2}\,\Delta^+$.
 A non-zero triplet vev {$v_\Delta^{} \equiv \sqrt{2}\,\langle \Delta^0 \rangle$
arises from the minimisation of the scalar potential and
leads to the following mass matrix for Majorana neutrinos:
%-------------------------------
\begin{equation}
%\mLll = 2\hll \langle\Delta^0\rangle = \sqrt{2}\hll v_{\Delta} .
\mLll = \sqrt{2}\hll v_{\Delta} .
\label{Eq:nu_massL}
\end{equation}
%-------------------------------

 The most general $\SU(2)_L\otimes \U(1)_Y$ invariant form
of the scalar potential is given
in Refs.~\cite{Cheng:1980qt,Ma:2000wp, Chun:2003ej}
and a detailed study of the theoretical constraints on its parameters 
has been performed in Ref.~\cite{Arhrib:2011uy}.
The  conservation of lepton number is broken by two units
due to a soft-breaking term $\mu \Phi^T i\sigma_2 \Delta^\dagger \Phi$
(here $\mu$ is a dimensional coupling constant),
which gives rise to $v_\Delta^{}$ and thus neutrino masses.
 This soft-breaking term
might be suppressed by a radiative mechanism~\cite{Kanemura:2012rj}.

The direct connection between $\hll$ and $\mLll$
in eq.~(\ref{Eq:nu_massL}) gives rise to phenomenological predictions
for processes which depend on $\hll$ (e.g.\ Ref.~\cite{Ma:2000wp})
because $\mLll$ has been severely restricted  
by neutrino oscillation measurements%
~\cite{Ref:solar-v, Ref:atm-v, Ref:acc-v,
Ref:MINOS-app, Ref:T2K-app,
Ref:CHOOZ, Ref:DayaBay, Ref:RENO, Ref:DoubleChooz, Abe:2013sxa,
Ref:long-reac-v}.
 One can write $\hll$ in terms of
the Maki-Nakagawa-Sakata~(MNS) matrix $U_\MNS$~\cite{Maki:1962mu}
and the diagonalised neutrino mass matrix as follows:
%-------------------------------
\begin{equation}
h_{\ell\ell^\prime}
=
 \frac{1}{ \sqrt{2}\, v_\Delta^{} } \mLll
=
 \frac{1}{ \sqrt{2}\, v_\Delta^{} }
 \Bigl[
  U_\MNS^\ast\,
  \text{diag}( m_1, m_2 e^{i\phi_1}, m_3 e^{i\phi_2} )
  U_\MNS^\dagger
 \Bigr]_{\ell\ell^\prime} ,
\label{Eq:hll}
\end{equation}
%-------------------------------
where $\phi_1$ and $\phi_2$ are the so-called Majorana phases.
 The MNS matrix is parametrised as
%-------------------------------
\begin{eqnarray}
U_\MNS
=
\begin{pmatrix}
 1 & 0 & 0\\
 0 & c_{23} & s_{23}\\
 0 & -s_{23} & c_{23}
\end{pmatrix}
\begin{pmatrix}
 c_{13} & 0 & s_{13} e^{-i\delta}\\
 0 & 1 & 0\\
 -s_{13} e^{i\delta} & 0 & c_{13}
\end{pmatrix}
\begin{pmatrix}
 c_{12} & s_{12} & 0\\
 -s_{12} & c_{12} & 0\\
 0 & 0 & 1
\end{pmatrix} ,
\label{Eq:MNS}
\end{eqnarray}
%-------------------------------
where $s_{ij}^{}$ and $c_{ij}^{}$
denote $\sin\theta_{ij}$ and $\cos\theta_{ij}$, respectively.

Clearly the decay widths of $H^{\pm\pm}\to \ell^\pm{\ell^\prime}^\pm$ depend on $\hll$
through eq.~(\ref{Eq:trip_yuk}).
The first quantitative studies of BR($H^{\pm\pm}\to \ell^\pm{\ell^\prime}^\pm$) in the HTM
were performed in Ref.~\cite{Chun:2003ej},
with further studies in Refs.%
~\cite{Akeroyd:2005gt, Ref:mnu-HTM-LHC, Perez:2008ha, Ref:multi-lepton-LHC}.
 Importantly,
$\BR(H^{\pm\pm}\to \ell^\pm{\ell^\prime}^\pm)$ depends on
the two unknown Majorana phases
and the absolute mass of the lightest neutrino
i.e.\ parameters which cannot be probed
in neutrino oscillation experiments. Thus
information on such parameters can be obtained
if $\BR(H^{\pm\pm}\to \ell^\pm{\ell^\prime}^\pm)$ are measured~\cite{Ref:mnu-HTM-LHC}.
A study on the relation
between BR($H^{\pm\pm}\to \ell^\pm{\ell^\prime}^\pm$)
and neutrinoless double beta decay
was performed in Ref.~\cite{Petcov:2009zr}.

 A distinctive signal of the HTM would be the observation of $H^{\pm\pm}$,
whose mass ($m_{H^{\pm\pm}}^{}$) may be of the order of the electroweak scale.
 Such particles could be produced with sizeable rates
at hadron colliders through the processes $\qqHppHmm$%
~\cite{Barger:1982cy, Gunion:1989in, Muhlleitner:2003me, Han:2007bk, Huitu:1996su}
and $\qqHpmpmHmp$~\cite{Barger:1982cy, Dion:1998pw, Akeroyd:2005gt}.
 Direct searches in these channels
have been carried out by the ATLAS~\cite{ATLAS:2012hi}
and CMS collaborations~\cite{Chatrchyan:2012ya},
using about $5\,\fb^{-1}$ of data at $\sqrt s=7\,\TeV$.
 The strongest limits are
for the channels $H^{\pm\pm}\to \ell^\pm{\ell^\prime}^\pm$
where $\ell,\ell^\prime$ is $e$ or $\mu$.
 For the case of $\BR(H^{\pm\pm}\to \ell^\pm{\ell^\prime}^\pm)=100\%$,
lower bounds of the order $m_{H^{\pm\pm}}^{} > 400\,\GeV$
have been derived.
 For $\BR(H^{\pm\pm}\to \ell^\pm{\ell^\prime}^\pm) \ll 100\%$
the mass limits are much weaker
e.g.\ $m_{H^{\pm\pm}}^{}>100\,\GeV$
for $\BR(H^{\pm\pm}\to \ell^\pm{\ell^\prime}^\pm)=1\%$.
 At present
there have been no direct searches for the decay mode
$H^{\pm\pm}\to W^\pm W^{\pm (*)}$,
which is the dominant decay for $v_{\Delta}> 1\,\MeV$.
However,
since $H^{\pm\pm}\to W^\pm W^{\pm (*)}$ would also give rise to
a multi-lepton signature with same-sign leptons,
the study in Ref.~\cite{Kanemura:2013vxa} applies
the selection cuts for a search for same-sign leptons
in Ref.~\cite{ATLAS:2012mn} 
to the case of $H^{\pm\pm}\to W^\pm W^{\pm (*)}$
and obtains the lower bound $m_{H^{\pm\pm}} > 60\,\GeV$.
 If $m_{H^{\pm\pm}}^{}> m_{H^{\pm}}^{}$
then the decay $H^{\pm\pm}\to H^\pm W^{\pm *}$ can be dominant,
even for relatively small mass splittings
$m_{H^{\pm\pm}}^{}- m_{H^{\pm}}^{}$.
 At present there has been no direct search in this channel.

 In this work we will study in detail
the branching ratios of the leptonic decays
of the singly charged Higgs,
$\BR_{\ell\nu} \equiv \sum_i \BR(H^\pm\to \ell^\pm\nu_i)$.
 We assume the scenario of $v_{\Delta}< 0.1\,\MeV$
for which $\sum_\ell \BR_{\ell\nu} \sim 1$,
with $\BR(H^\pm\to tb)$ and $\BR(H^\pm \to WZ)$
(which are $\propto v_\Delta^2$)
being negligible (see e.g.\ Ref.~\cite{Perez:2008ha}).
 In order to avoid decays of the form
$H^\pm \to H^0 W^*$%
~\cite{Gunion:1998ii, Chakrabarti:1998qy,
Chun:2003ej, Akeroyd:2005gt, Perez:2008ha} and
$H^\pm \to H^{\pm\pm}W^*$~\cite{Akeroyd:2011zza}
(which can be dominant in the HTM)
we assume $m_{H^\pm}^{} \simeq m_{H^0}^{} \simeq m_{H^{\pm\pm}}^{}$.
 Since the vertex $H^\pm tb$ is suppressed by $v_{\Delta}$
the decay width for $t\to H^\pm b$
with $m_{H^\pm}^{} < m_t - m_b$
is negligible,
and thus searches at the LHC in this channel will have no sensitivity.
 There were searches at the CERN LEP experiment for
$e^+e^-\to H^+H^-$ with $H^\pm \to \tau^\pm\nu$,
in which the limit $m_{H^\pm} \gsim 90\,\GeV$ was derived~\cite{Abbiendi:2013hk}.
For the decay channels $H^\pm \to e^\pm\nu$ and $H^\pm \to \mu^\pm\nu$,
the limits from explicit searches at LEP for 
sleptons $\tilde{\ell}$
in supersymmetric models can be applied~\cite{Abbiendi:2003ji}
(i.e.\
searches for $e^+e^-\to \tilde\ell^+\tilde\ell^-$
with $\tilde{\ell}^\pm\to \ell^\pm\chi^0_1$
for $\ell^\pm=e^\pm,\mu^\pm$, where
$\chi^0_1$ is the lightest neutralino which appears as missing energy).
Again, these limits can be satisfied by $m_{H^\pm} \gsim 90\,\GeV$.

Previous studies of $\BR_{\ell\nu}$ in the HTM
have been performed in Ref.~\cite{Perez:2008ha}.
 The partial width of $H^{\pm}\to \ell^\pm \nu_i$
is determined from eq.~(\ref{Eq:trip_yuk})
and is proportional to $|(U_\MNS^T h)_{i \ell}|^2$.
 After summing over the three mass eigenstates of neutrinos,
the summed partial width $\Gamma(H^{\pm}\to \ell^\pm \nu)$ is given by
%-------------------------------
\begin{eqnarray}
\Gamma(H^\pm \to \ell^\pm \nu)
=
 \frac{ m_{H^\pm}^{} }{8\pi}
 \left( h^\dagger h \right)_{\ell\ell}
=
 \frac{ m_{H^\pm}^{} }{16\pi v_\Delta^2}
 \left( m_L^\dagger m_L^{} \right)_{\ell\ell} .
\end{eqnarray}
%-------------------------------
Note that the summation ensures that
the dependence on the Majorana phases vanishes,
unlike the case for $\Gamma(H^{\pm\pm}\to \ell^\pm\ell^\pm$),
and this notable result was first pointed out in Ref.~\cite{Perez:2008ha}.
Explicit forms of $(m_L^\dagger m_L^{})_{\ell\ell}$
are given by
%-------------------------------
\begin{eqnarray}
(m_L^\dagger m_L^{})_{ee}
&=&
 m_1^2
 + s_{13}^2 \Delta m^2_{31}
 + s_{12}^2 c_{13}^2 \Delta m^2_{21}
\label{Eq:hh_ee_N}
\\
&=&
 m_3^2
 + c_{13}^2 \Delta m^2_{13}
 + s_{12}^2 c_{13}^2 \Delta m^2_{21} ,
\label{Eq:hh_ee_I}
\\
%
%%%%%%%%%%%%%%%%%%%%%%%%%
(m_L^\dagger m_L^{})_{\mu\mu}
&=&
 m_1^2
 + s_{23}^2 c_{13}^2 \Delta m^2_{31}
 + ( c_{12}^2 c_{23}^2 + s_{12}^2 s_{23}^2 s_{13}^2 ) \Delta m^2_{21}
\nonumber\\
&&\hspace*{60mm}
{}- 2 c_{12} s_{12} c_{23} s_{23} s_{13} \Delta m^2_{21} \cos\delta
\label{Eq:hh_mm_N}
\\
&=&
 m_3^2
 + ( 1 - s_{23}^2 c_{13}^2 ) \Delta m^2_{13}
 + ( c_{12}^2 c_{23}^2 + s_{12}^2 s_{23}^2 s_{13}^2 ) \Delta m^2_{21}
\nonumber\\
&&\hspace*{60mm}
{}- 2 c_{12} s_{12} c_{23} s_{23} s_{13} \Delta m^2_{21} \cos\delta ,
\label{Eq:hh_mm_I}
\\
%
%%%%%%%%%%%%%%%%%%%%%%%%%
(m_L^\dagger m_L^{})_{\tau\tau}
&=&
 m_1^2
 + c_{23}^2 c_{13}^2 \Delta m^2_{31}
 + ( c_{12}^2 s_{23}^2 + s_{12}^2 c_{23}^2 s_{13}^2 ) \Delta m^2_{21}
\nonumber\\
&&\hspace*{60mm}
{}+ 2 c_{12} s_{12} c_{23} s_{23} s_{13} \Delta m^2_{21} \cos\delta
\label{Eq:hh_tt_N}
\\
&=&
 m_3^2
 + ( 1 - c_{23}^2 c_{13}^2 ) \Delta m^2_{13}
 + ( c_{12}^2 s_{23}^2 + s_{12}^2 c_{23}^2 s_{13}^2 ) \Delta m^2_{21}
\nonumber\\
&&\hspace*{60mm}
{}+ 2 c_{12} s_{12} c_{23} s_{23} s_{13} \Delta m^2_{21} \cos\delta ,
\label{Eq:hh_tt_I}
\end{eqnarray}
%-------------------------------
where $\Delta m^2_{ij} \equiv m_i^2 - m_j^2$.
The effect of the CP-violating phase $\delta$ is negligible \cite{Xing:2013woa}
because it appears with $s_{13}^{} \Delta m^2_{21}$,
which is much smaller than $|\Delta m^2_{31}|$
(with an ${\mathcal O}(1)$ coefficient)
in the second term in the right hand-side
of eqs.~(\ref{Eq:hh_mm_N})-(\ref{Eq:hh_tt_I}).

In the HTM the production process $q'\overline q\to W\to H^{\pm\pm}H^{\mp}$
affords the best detection prospects for $H^\pm\to \ell^\pm\nu$
for a given $m_{H^\pm}^{}$. This mode (with $m_{H^{\pm\pm}}^{} = m_{H^\pm}^{}$)
has already been taken into account in the search for $H^{\pm\pm}$ by the 
CMS collaboration in Ref.~\cite{Chatrchyan:2012ya}. To our knowledge
there has not been a dedicated search for
$q\overline q\to \gamma, Z\to H^+H^-$ at the LHC,
and we are not aware of a simulation of the detection prospects
for $\sqrt 8\,\TeV$ and ${ \cal L}\simeq 20\,\fb^{-1}$.

%%%%%%%%%%%%%%%%%%%%%%%%
%%%  subsec: nu2HDM  %%%
%%%%%%%%%%%%%%%%%%%%%%%%
\subsection{$\nu$2HDM}
 In the $\nu$2HDM,
the SM is extended with three right-handed gauge singlet fermions $\nu_{iR}^{}$
and a second scalar $\SU(2)_L$-doublet $\Phi_\nu =(\phi_\nu^+, \phi_\nu^0)^T$,
which is in the same representation as $\Phi$ under the SM gauge group.
 A global U(1) symmetry is imposed,
under which $\Phi_\nu$ and the three $\nu_{iR}^{}$ have charge $+1$
and all the other fields are uncharged~\cite{Davidson:2009ha}.
The following Yukawa interaction is added to that of the SM:
%-------------------------------
\begin{equation}
{\mathcal L}^{\text{$\nu$2HDM}}_{\text{Yuk}}
=
 (y_\nu^{})_{i \ell}\,
 \overline{ \nu_{iR}^{} }\, \Phi_\nu^T\, i\sigma_2\, L_\ell
 + {\rm h.c.} ,
\label{eq:smyuk}
\end{equation}
%-------------------------------
where $(y_\nu^{})_{i \ell}$ is the $3\times 3$ matrix
of Yukawa coupling constants for neutrinos.
Note that the U(1) symmetry forbids Majorana mass terms
$\frac{1}{\,2\,} m_{iR}^{} (\nu_{iR}^{})^T C\, \nu_{iR}^{}$.
 If the global U(1) symmetry is softly broken only by $m_{12}^2 \Phi^\dagger \Phi_\nu$%
~\cite{Davidson:2009ha},
there arises a vev $v_\nu^{} \equiv \sqrt{2}\, \langle \phi_\nu^0 \rangle$ 
and lepton number is conserved.\footnote{Since the Majorana mass terms of the right-handed neutrinos
also softly break the global $U(1)$ symmetry,
it may be better to impose by hand
the conservation of the lepton number  on the Lagrangian.}
The smallness of the neutrino masses can be naturally understood
if the soft-breaking term is generated
at the loop level~\cite{Chang:1986bp, Kanemura:2013qva}.}
Then, $\nu_{\ell L}^{}$ and $\nu_{i R}^{}$ become three Dirac neutrinos
whose mass matrix is simply given by
%-------------------------------
\begin{eqnarray}
\mDil = (y_\nu^{})_{i\ell}\, \frac{v_\nu^{}}{\sqrt{2}} .
\label{Eq:nu_massD}
\end{eqnarray}
%-------------------------------
One can take $\nu_{iR}^{}$ as the right-handed components
of the mass eigenstates $\nu_i^{}$ without loss of generality, which leads to 
the following expression:
%-------------------------------
\begin{eqnarray}
(y_\nu^{})_{i \ell}
=
 \frac{\sqrt{2}\,}{v_\nu^{}} \mDil
=
 \frac{\sqrt{2}\,}{v_\nu^{}}
 \Bigl[
  \text{diag}( m_1, m_2, m_3 ) U_\MNS^\dagger
 \Bigr]_{i \ell} .
\label{Eq:yil}
\end{eqnarray}
%-------------------------------

 The charged Higgs $H^-$ in the $\nu$2HDM
decays into $\ell_L \overline{\nu_R^{}}$
while $H^-$ in the HTM decays into $\ell_L \nu_L^{}$.
The partial decay widths for $H^\pm \to \ell^\pm \nu$
(summed over all the neutrino species) are calculated as
%-------------------------------
\begin{eqnarray}
\Gamma(H^\pm \to \ell^\pm \nu)
=
 \frac{m_{H^\pm}^{}}{16\pi}
 \left( y_\nu^\dagger y_\nu^{} \right)_{\ell\ell}
=
 \frac{m_{H^\pm}^{}}{8\pi v_\nu^2} 
 \left( m_D^\dagger m_D^{} \right)_{\ell\ell} .
\end{eqnarray}
%-------------------------------
 It is evident that
$(m_D^\dagger m_D^{})_{\ell\ell^\prime}
= (m_L^\dagger m_L^{})_{\ell\ell^\prime}$
with eqs.~\eqref{Eq:hll} and \eqref{Eq:yil},
and explicit expressions are presented
in eqs.~\eqref{Eq:hh_ee_N}-\eqref{Eq:hh_tt_I}. Thus,
the dependence of $\Gamma(H^\pm \to \ell^\pm \nu)$ on the neutrino parameters
is identical in both the HTM and the $\nu$2HDM\@.
Previous studies of $\BR_{\ell\nu}$ in the $\nu$2HDM
have been performed in Ref.~\cite{Davidson:2009ha}.
 Detection prospect of $H^\pm$ at LHC
is discussed in Ref.~\cite{Davidson:2010sf}.

 In the $\nu$2HDM,
the cross section for $q\overline q\to \gamma, Z\to H^+H^-$
is larger than that in the HTM by a factor of 2.7.
This is a consequence of the different isospin of $H^\pm$~$(I_3=0)$ in the HTM
and $H^\pm$~$(I_3=\pm 1/2)$ in the $\nu$2HDM\@.
Hence the detection prospects in the channel $q\overline q\to \gamma, Z\to H^+H^-$
are significantly better in the $\nu$2HDM than in the HTM,
as emphasised in Ref.~\cite{Davidson:2010sf}.

\subsection{Neutrino oscillation parameters}

As shown above, the decay widths of $H^\pm\to \ell^\pm\nu$
depend on the neutrino parameters.
Neutrino oscillation experiments involving solar~\cite{Ref:solar-v}, 
atmospheric~\cite{Ref:atm-v},
accelerator~\cite{Ref:acc-v, Ref:MINOS-app, Ref:T2K-app},
and reactor neutrinos%
~\cite{Ref:CHOOZ, Ref:DayaBay, Ref:RENO, Ref:DoubleChooz, Abe:2013sxa,
Ref:long-reac-v}
are sensitive to the mass-squared differences and the mixing angles, 
and give the following preferred values and ranges:
%-------------------------------
\begin{eqnarray}
&&
\Delta m^2_{21} \simeq 7.5\times 10^{-5} {\rm eV}^2 \,,~~
|\Delta m^2_{31}| \simeq 2.3\times 10^{-3} {\rm eV}^2\,, \\
&&
\sin^2{2\theta_{12}}\simeq 0.85 \,,~~~~
0.4 \lesssim s_{23}^2 \lesssim 0.6 \,,~~~~
0.07 \lesssim \sin^2{2\theta_{13}} \lesssim 0.11\,,~~~~~~~~~~~~
\label{Eq:obs_para}
\end{eqnarray}
%-------------------------------
where $\theta_{12}, \theta_{13} < \pi/4$.
We use these values in our numerical analysis unless otherwise mentioned.
 Varying $|\Delta m^2_{31}|$, $\Delta m^2_{21}$ and  $\sin^2{2\theta_{12}}$
within their allowed ranges only causes a very small error in $\BR_{\ell\nu}$
e.g.\ the value of $\BR_{e\nu}$ in our analysis
with $m_1 \simeq 0$~($\Delta m^2_{31} > 0$)
has about a $10\,\%$ error in total from varying them,
while the effect of varying
$\sin^2{2\theta_{13}}$ (which we will study in detail) is much larger.
Information on the mass $m_0$ of the lightest neutrino
and the Majorana phases
cannot be obtained from neutrino oscillation experiments.
 This is because
the oscillation probabilities are independent of these parameters,
not only in vacuum but also in matter.
 If $m_0 \gtrsim 0.2\,\eV$,
a future ${}^3$H beta decay experiment~\cite{Ref:KATRIN}
can measure $m_0$.
 Experiments which seek neutrinoless double beta decay
(See e.g., Ref.~\cite{Avignone:2007fu} for a review)
are only sensitive to a combination of neutrino masses and phases
when neutrinos are Majorana fermions.

 The value of $\delta$ in completely unknown.
 Measurement of $\delta$ is a main goal
of oscillation experiments with accelerator neutrinos%
~\cite{Ref:LOI-JHF, Abe:2011ts, Ayres:2004js, Choubey:2011zzq}.
 The measurement uses appearance modes~(e.g.\ $\nu_\mu^{} \to \nu_e^{}$)
whose dominant terms are controlled by
$s_{23}^2$ and $\sin^2(2\theta_{13})$.
 Since this CP-violating parameter is extracted 
by comparing measurements
with a neutrino beam and an antineutrino beam~(not on the anti-Earth),
the measurement of $\delta$ is affected
by the sign of $\Delta m^2_{31}$ due to the effect of the Earth's matter on the oscillations.

 Since the sign of $\Delta m_{31}^2$ is also undetermined at present, 
distinct neutrino mass spectrums are possible.
 The case with $\Delta m^2_{31} >0$ is referred to as
{\it Normal mass ordering}~(NO) where $m_1 < m_2 < m_3$
and the case with $\Delta m^2_{31} <0$ is known as
{\it Inverted mass ordering}~(IO) where $m_3 <  m_1 < m_2$.
The sign of $\Delta m_{31}^2$ can be determined by 
long baseline oscillation measurements%
~(e.g.\ in the NOvA experiment~\cite{Ayres:2004js})
and precise measurements of the oscillations of atmospheric neutrinos%
~(e.g.\ with the Hyper-Kamiokande~\cite{Abe:2011ts}).

An important recent result is the knowledge that
the small mixing angle $\theta_{13}$ 
is now known to be significantly different from zero.
 The nonzero value of $\theta_{13}$ 
makes the measurement of leptonic CP-violation possible 
(which depends on $s_{13}^{} \sin\delta$,
as can be seen from Eq.~\eqref{Eq:MNS}) at future 
experiments.
 Reactor experiments probe the probability of the disappearance of
anti-electron neutrinos~($\overline\nu_e$), 
a process which is sensitive to $\sin^2{2\theta_{13}}$.
 The Daya Bay collaboration has obtained the value
$\sin^2{2\theta_{13}}=0.089\pm 0.010\pm 0.005$~\cite{Ref:DayaBay};
 the RENO collaboration has obtained
$\sin^2{2\theta_{13}}=0.113\pm 0.013\pm 0.014$~\cite{Ref:RENO}; 
 the Double Chooz collaboration has obtained
$\sin^2{2\theta_{13}}=0.109\pm 0.030\pm 0.025$
(with a Gadolinium analysis)~\cite{Ref:DoubleChooz} and
$\sin^2{2\theta_{13}}=0.097\pm 0.034\pm 0.034$ 
(with an analysis which captures neutrons on hydrogen)~\cite{Abe:2013sxa}.
 Long baseline experiments search for
the appearance of $\nu_e$ from a beam of $\nu_\mu$,
and this process is sensitive to
the combination $s_{23}^2 \sin^2{2\theta_{13}}$.
 Assuming $\theta_{23}=\pi/4$,
T2K has obtained
$\sin^2{2\theta_{13}}=0.088^{+0.049}_{-0.034}$~\cite{Ref:T2K-app}
(See also a preliminary update~\cite{Ref:T2K-EPS2013}).
The NOvA experiment~\cite{Patterson:2012zs}
will also measure $s_{23}^2 \sin^2{2\theta_{13}}$.
The ultimate precision is expected to be about
$0.005$ at 68\% confidence level~(c.l.)
at the Daya Bay~(see e.g.\ Ref.~\cite{Qian:2012df}).

 The mixing angle $\sin\theta_{23}$ is known to be almost maximal.
 The currently preferred $2\sigma$ range is
$0.4 \lesssim s_{23}^2 \lesssim 0.6$~\cite{Ref:atm-v}.
 Long baseline experiments~\cite{Ref:LOI-JHF, Ayres:2004js}
will further improve the precision in the determination of $s_{23}^2$
by studying the survival probability of $\nu_\mu$,
which is proportional to $\sin^2{2\theta_{23}}$.
 However,
if $\theta_{23}$ deviates enough from $\pi/4$,
there are two possible values of $\theta_{23}$
which give the same value of $\sin^2{2\theta_{23}}$.
 For example,
$s_{23}^2 = 0.4$ and $0.6$ are obtained
for $\sin^2{2\theta_{23}} = 0.96$;
 this gives about $\pm 20\%$ uncertainty in appearance probabilities
(e.g.\ $\nu_\mu \to \nu_e$)
which will be used to measure $\delta$.
 The ambiguity (the "octant degeneracy") on whether $s_{23}^2 > 0.5$ or not
can be resolved by precise measurement
of the atmospheric neutrino%
~(e.g.\ with the Hyper-Kamiokande experiment~\cite{Abe:2011ts}).
 It is also possible to resolve the ambiguity
by e.g.\ utilising the complementarity
of reactor and long baseline experiments~\cite{Ref:reac-LBL}.

%%%%%%%%%%%%%%%%%%%%%%%%%
%%%%  sec: pheno  %%%%%%%
%%%%%%%%%%%%%%%%%%%%%%%%%
\section{BR$(H^\pm\to \ell^\pm\nu$) in the HTM and in the neutrinophilic 2HDM}

 The dependence of
$\BR_{\ell\nu}$ on the neutrino parameters has been studied in Ref.~\cite{Perez:2008ha}
in the context of the HTM,
and in Ref.~\cite{Davidson:2009ha}
in the context of the $\nu$2HDM\@.
 Both studies are in agreement,
and present $\BR_{\ell\nu}$
as functions of the lightest neutrino mass~($m_0$)
for both orderings of neutrino masses.
 In Refs.~\cite{Perez:2008ha,Davidson:2009ha}
the BRs of the leptonic decay channels of $H^\pm$
were displayed as scatter plots
in which the neutrino mixing angles and mass differences
were varied over the allowed intervals.
 From those studies
it is not clear which values of $s_{23}^2$ and $\sin^2{2\theta_{13}}$
correspond to the upper and lower limits of the allowed regions of the BRs.
 Correlations of BRs with respect to the neutrino parameters are not also clear.
 In this work
we clarify the effect of varying
$s_{23}^2$ and $\sin^2{2\theta_{13}}$ individually,
and quantify the impact of the recent measurement of $\sin^2{2\theta_{13}}$
on the leptonic BRs of $H^\pm$.
 Where comparison is possible our results are in agreement with
those in Refs.~\cite{Perez:2008ha,Davidson:2009ha}.
 As explained earlier,
we only consider the parameter space of
$v_\Delta^{} (v_\nu^{}) < 0.1\,\MeV$
for which $\sum_\ell \BR_{\ell\nu} \sim 1$.
 Thus the dependence of $\BR_{\ell\nu}$ on
$m_{H^\pm}^{}$ drops out,
but for definiteness we fix
$m_{H^\pm}=500\,\GeV$~($200\,\GeV$)
in the HTM~($\nu$2HDM).
 Since we assume $m_{H^\pm}=m_{H^{\pm\pm}}$ in the HTM,
the choice of $m_{H^\pm}=500\,\GeV$ is necessary in order 
to comfortably satisfy  current limits on 
$m_{H^{\pm\pm}}$ from direct searches for $H^{\pm\pm}\to \ell^\pm\ell^\pm$
at the LHC~\cite{ATLAS:2012hi,Chatrchyan:2012ya}.
 We also fix $v_\Delta^{} (v_\nu^{}) =1000\,\eV$.
 Although
$\BR_{\ell\nu}$ is not sensitive to the exact value of $v_\Delta^{} (v_\nu^{})$ for $v_\Delta^{} (v_\nu^{}) < 0.1\,\MeV$,
contributions of the scalars $H^{\pm\pm}$~(in the HTM) and $H^{\pm}$
to lepton-flavour-violating decays
such as $\mu\to eee$ and $\tau\to \ell\ell\ell$~(in the HTM),
and $\mu\to e\gamma$ are sensitive to the value%
~\cite{Davidson:2009ha,Chun:2003ej,Akeroyd:2009nu}.
 Constraints from these decays are satisfied
for $v_\Delta^{} \gsim 1000\,\eV$,
and so we fix $v_\Delta^{} (v_\nu^{}) =1000\,\eV$.

%-------------------------
\begin{figure}[t]
\begin{center}
\includegraphics[origin=c, angle=-90, scale=0.6]{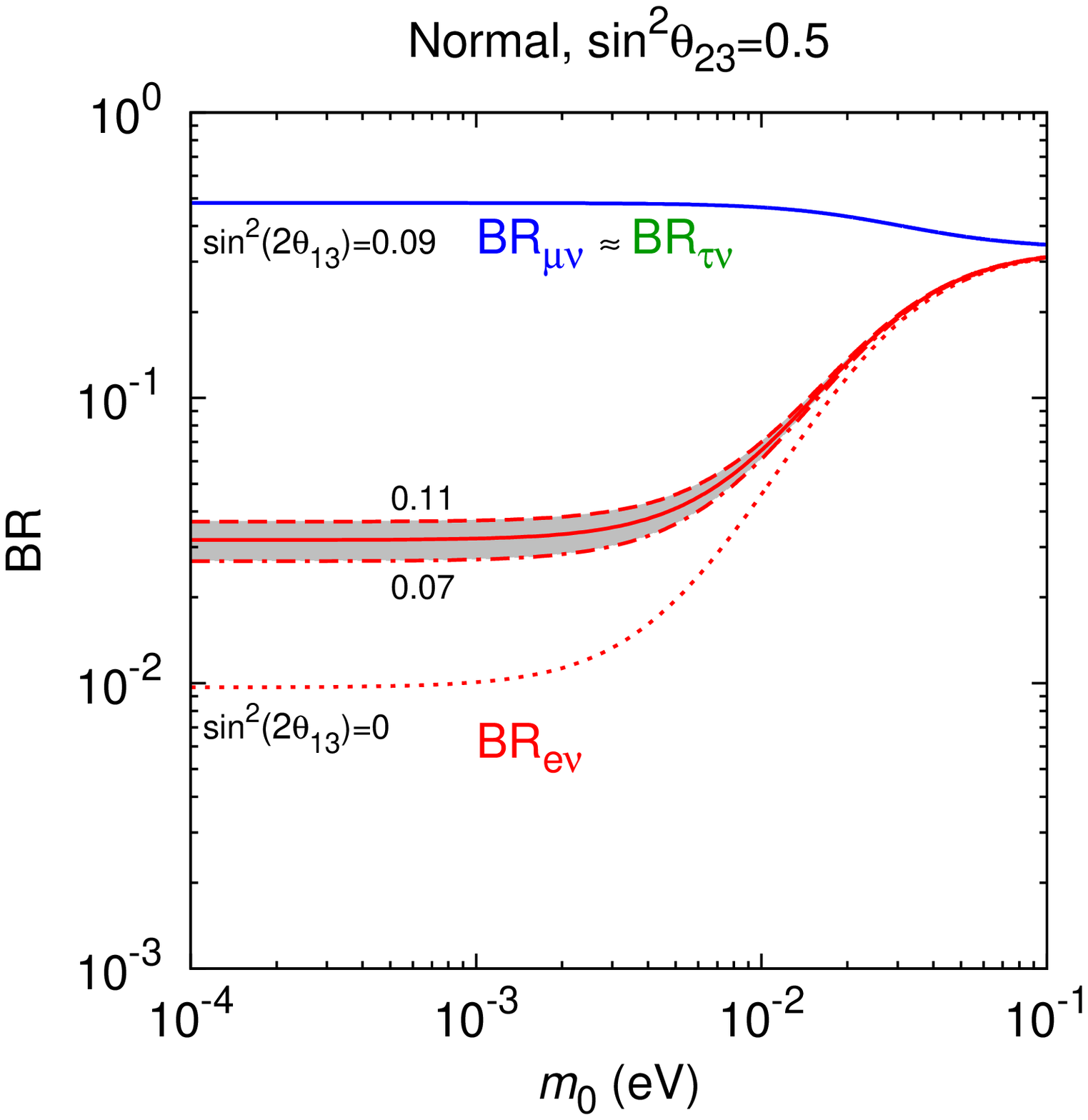}
\\\vspace*{-20mm}
\hspace*{2mm}
\includegraphics[origin=c, angle=-90, width=93.5mm, height=100mm]{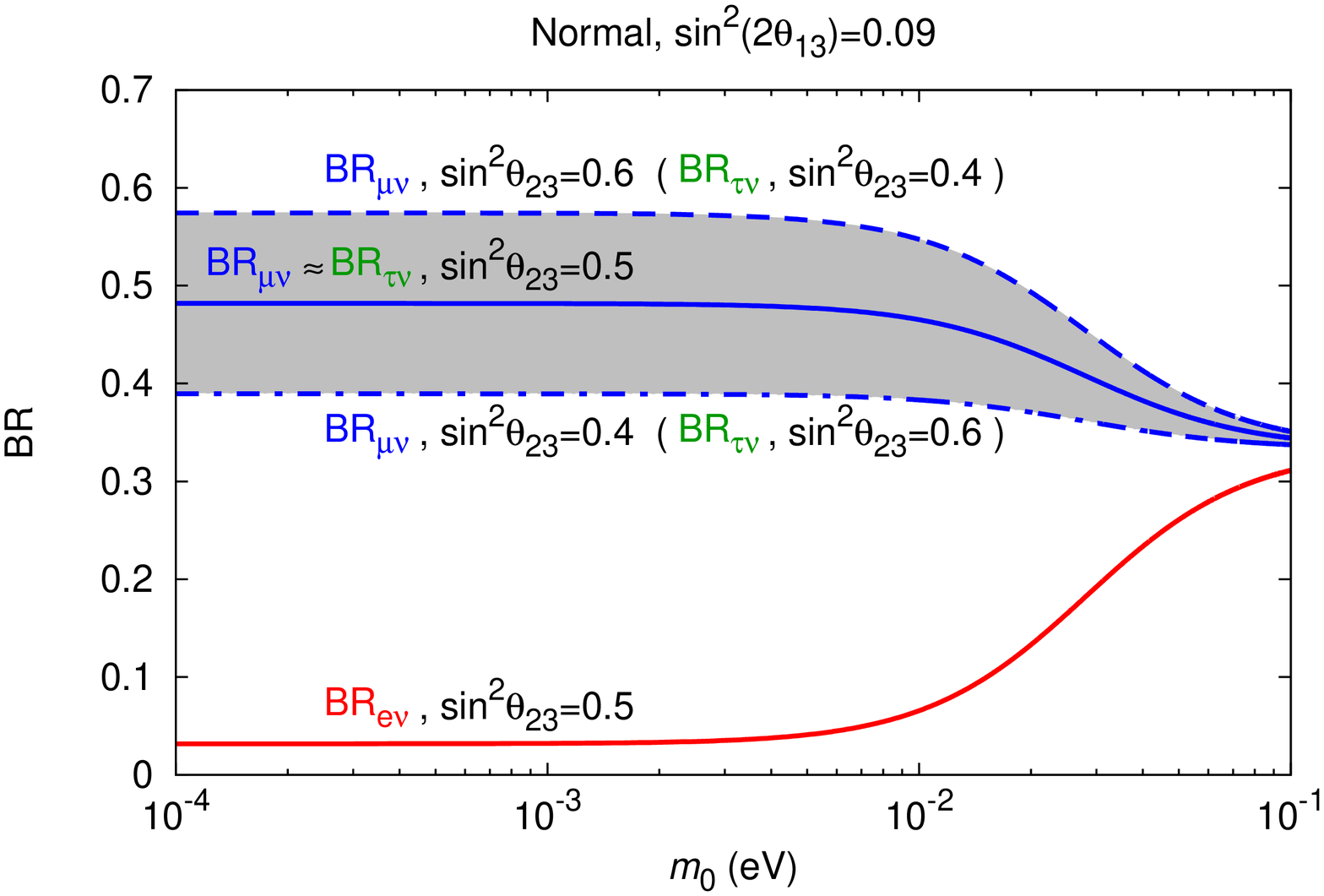}
\vspace*{-20mm}
\caption{
 BR($H^\pm\to e^\pm\nu$), BR($H^\pm\to \mu^\pm\nu$) and 
BR($H^\pm\to \tau^\pm\nu$) are plotted as functions of
$m_0$~($=m_1$) for the case of a normal mass ordering.
 Upper panel: for three different values of
$\sin^2{2\theta_{13}}$ and fixing  $s_{23}^2=0.5$.
 Solid lines are obtained for $\sin^2{2\theta_{13}} = 0.09$.
 Lower panel: for three different values of
$s_{23}^2$ and fixing $\sin^2{2\theta_{13}}=0.09$.
}
\label{Fig:BRs-N}
\end{center}
\end{figure}
%-------------------------

%%%%%%%  Normal mass ordering, s_13  %%%%%%%%
 In Fig.~\ref{Fig:BRs-N}~(upper panel)
for the normal mass ordering where $m_0 = m_1$,
we plot $\BR_{e\nu}$ with light (red) lines
and $\BR_{\mu\nu}$ with a dark (blue) line
as functions of $m_0$
for four different values of $\sin^2{2\theta_{13}}$:
%-------------------------
\begin{align*}
& \sin^2{2\theta_{13}}=0.11 \
&
& \hspace*{-10mm}
  \text{(the upper limit at about 95\% c.l., a dashed line)} ,\\
%
%-----------
& \sin^2{2\theta_{13}}=0.09 \
&
& \hspace*{-10mm}
  \text{(the experimental central value, solid lines)} ,\\
%
%-----------
& \sin^2{2\theta_{13}}=0.07 \
&
& \hspace*{-10mm}
  \text{(the lower limit at about 95\% c.l., a dot-dashed line)} ,\\
%
%-----------
& \sin^2{2\theta_{13}}=0 \
&
& \hspace*{-10mm}
  \text{(which is now excluded, a dotted line)} .
\end{align*}
%-------------------------

The values $\sin^2{2\theta_{13}}=0.11$, $0.09$, and $0.07$
correspond to $s_{13}^2 = 0.028$, $0.023$, and $0.018$, respectively.
 The maximal mixing $s_{23}^2 = 0.5$ is used,
and all other neutrino parameters are fixed as in eq.~\eqref{Eq:obs_para}.
 For this choice of $s_{23}^2$,
one has $\BR_{\mu\nu} \approx \BR_{\tau\nu}$,
a result which is due to an approximate $\mu$-$\tau$ exchange symmetry
of $U_\MNS$.
 Since $\BR_{\mu\nu}$ and $\BR_{\tau\nu}$
are not very sensitive to $\sin^2{2\theta_{13}}$,
we used only $\sin^2{2\theta_{13}} = 0.09$ for $\BR_{\mu\nu}$.
 They are much more sensitive to $s_{23}$.
 For $m_0 < 10^{-2}\,\eV$
it can be seen that $\BR_{e\nu}$
is very sensitive to the value of $\sin^2{2\theta_{13}}$.
 This can be understood from the explicit expression
for $(m_L^\dagger m_L^{})_{ee}$ in eq.~\eqref{Eq:hh_ee_N},
in which the term $s_{13}^2 \Delta m^2_{31}$ can be the dominant one
for $s_{13}^2 \gtrsim 0.01$ and $m_1 \lesssim 5\times 10^{-3}\,\eV$.

 The shaded region in Fig.~\ref{Fig:BRs-N}~(upper panel)
between the curves for $\sin^2{2\theta_{13}}=0.07$ and $0.11$
corresponds to the allowed region of $\BR_{e\nu}$ at about 95\%~c.l.
The lowest value is $\BR_{e\nu} \simeq 2.7\%$,
and is obtained for $\sin^2{\theta_{13}}=0.07$ and $m_0<10^{-3}\,\eV$.
 It is notable
that this minimum $\BR_{e\nu}$ is considerably larger than
the value $\BR_{e\nu}=1\%$
which is obtained for the (now strongly disfavoured) case
of $\sin^2{2\theta_{13}}=0$.
 Hence the measurement of $\sin^2{2\theta_{13}}$
has now disfavoured the parameter space of
$1\% < \BR_{e\nu} < 2.7\%$, and the minimum value of
$\BR_{e\nu}$ is now three times larger than before
for the case of the normal mass ordering.
 This result improves the detection prospects
of the channel $H^\pm\to e^\pm\nu$ at the LHC,
and will be discussed in more detail below\@.

%%%%%%%  Normal mass ordering, s_23  %%%%%%%%
 In Fig.~\ref{Fig:BRs-N}~(lower panel)
we show the $m_0$-dependence of
$\BR_{e\nu}$ with a light (red) line
and $\BR_{\mu\nu}$ with dark (blue) lines
for the case of the normal mass ordering,
but this time we fix $\sin^2{2\theta_{13}}=0.09$ and
consider three different values of $s_{23}^2$:
%-------------------------
\begin{align*}
& s^2_{23}=0.6 \
&
& \hspace*{-10mm}
  \text{(the upper limit at about 95\% c.l., a dashed line)} , \\
%
%------------
& s^2_{23}=0.5 \
&
& \hspace*{-10mm}
  \text{(maximal mixing, solid lines)} , \\
%
%------------
& s^2_{23}=0.4 \
&
& \hspace*{-10mm}
  \text{(the lower limit at about 95\% c.l., a dot-dashed line)} .
\end{align*}
%-------------------------
 Note that $s_{23}^2$ does not appear
in the expression for $(m_L^\dagger m_L^{})_{ee}$ in eq.~\eqref{Eq:hh_ee_N}
and so $\BR_{e\nu}$ is completely insensitive
to the value of $s_{23}^2$.
 Therefore in Fig~.\ref{Fig:BRs-N}~(lower panel)
we plot $\BR_{e\nu}$ for $s_{23}^2=0.5$ only.
 In contrast,
$\BR_{\mu\nu}$ and $\BR_{\tau\nu}$
are quite sensitive to $s_{23}^2$ e.g.\ for $m_0 < 10^{-2}\,\eV$,
where $\BR_{\mu\nu}$ takes the values
$\simeq 57\%$, {$\simeq$~$48\%$
and $\simeq$~$39\%$
for $s_{23}^2$~$=0.6$, $0.5$ and $0.4$,
respectively.
 Note that
$\BR_{\tau\nu}$ for $s^2_{23}=0.6$, $0.5$ and $0.4$
are almost given by
dot-dashed, solid and dashed curves of $\BR_{\mu\nu}$,
respectively.
 The case of $s_{23}^2>0.5$ leads to $\BR_{\mu\nu} > \BR_{\tau\nu}$,
while $s_{23}^2<0.5$ leads to $\BR_{\mu\nu} < \BR_{\tau\nu}$.
 If $\BR_{\mu\nu} > 48\,\%$ is measured then this would require
$s_{23}^2 > 0.5$ and $m_0 = m_1 \lesssim 10^{-2}\,\eV$.

%--------------------------------
\begin{figure}[t]
\begin{center}
\includegraphics[origin=c, angle=-90, width=93.5mm, height=100mm]
                {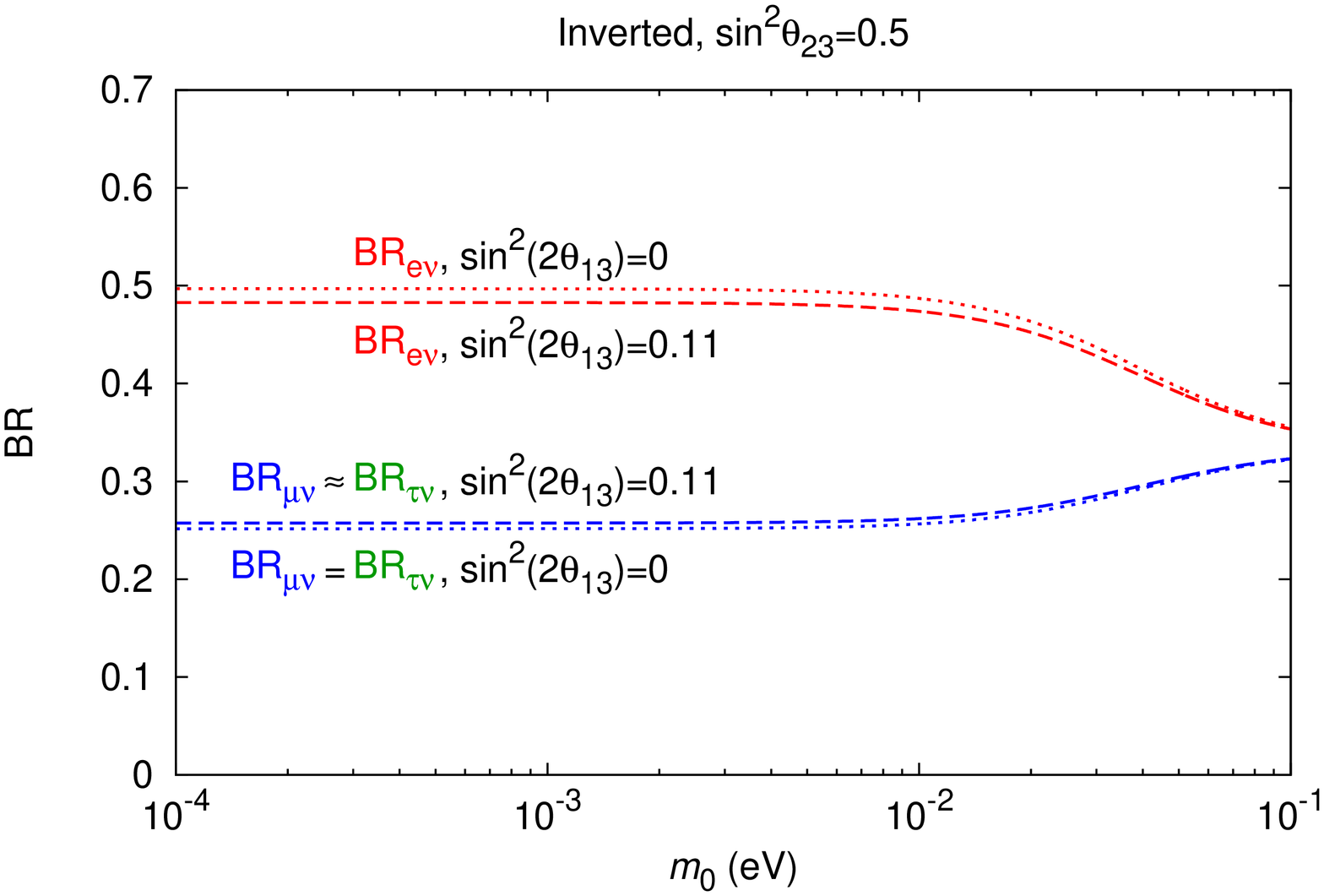}
\\\vspace*{-15mm}
\includegraphics[origin=c, angle=-90, width=93.5mm, height=100mm]
                {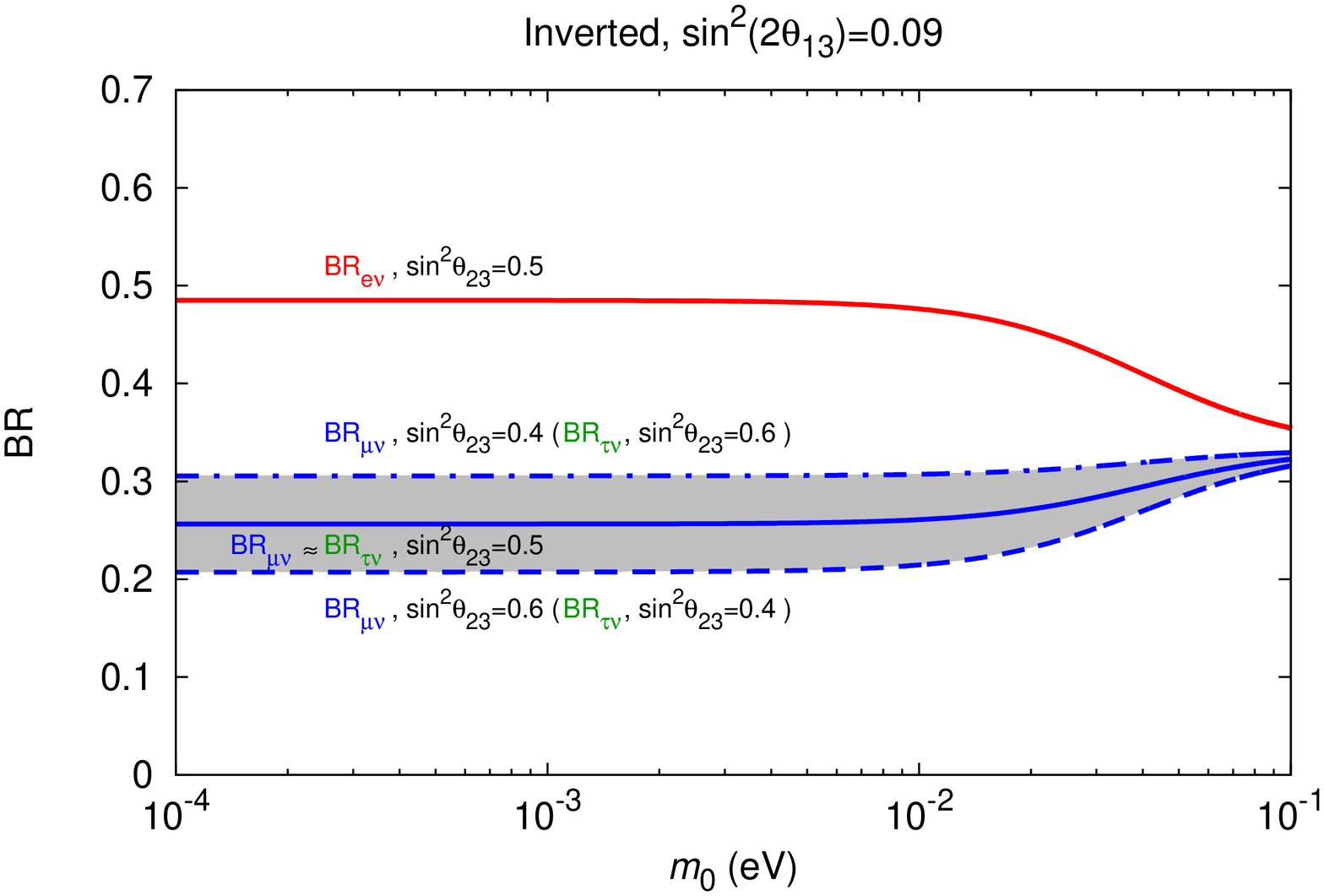}
\vspace*{-20mm}
\caption{
 BR($H^\pm\to e^\pm\nu$), BR($H^\pm\to \mu^\pm\nu$) and 
BR($H^\pm\to \tau^\pm\nu$) are plotted as functions of
$m_0$~($=m_3$) for the case of inverted mass ordering.
Upper panel: for three different values of
$\sin^2{2\theta_{13}}$ and fixing $s_{23}^2=0.5$.
Lower panel: for three different values of
$s_{23}^2$ and fixing $\sin^2{2\theta_{13}}=0.09$.
}
\label{Fig:BRs-I}
\end{center}
\end{figure}
%--------------------------------

%%%%%%%  Inverted mass ordering, s_13  %%%%%%%%
 We now discuss the case of the inverted mass ordering
where $m_0 = m_3$. Figure~\ref{Fig:BRs-I}~(upper panel)
is the analogue of Fig~.\ref{Fig:BRs-N}~(upper panel),
and considers only two values of $\sin^2{2\theta_{13}}$:
$\sin^2{2\theta_{13}}=0.11$
(approximately the 95\% c.l.\ upper limit, a dashed line)
and $\sin^2{2\theta_{13}}=0$
(which is now excluded, a dotted line).
 Since the dominant contribution of $\theta_{13}$ to the BRs
comes from the combination $c_{13}^2 \Delta m^2_{13}$
in eqs.~\eqref{Eq:hh_ee_I}, \eqref{Eq:hh_mm_I}, and \eqref{Eq:hh_tt_I},
one has the result that the BRs deviate by only a couple of percent
when $\sin^2{2\theta_{13}}$ is varied.
 Figure~\ref{Fig:BRs-I}~(lower panel) is
the analogue of Fig.~\ref{Fig:BRs-N}~(lower panel),
again fixing $\sin^2{2\theta_{13}}=0.09$ and
considering three different values of $s_{23}^2$ (=$0.4$, $0.5$ and $0.6$).
 Again one sees that the difference
between $\BR_{\mu\nu}$ and $\BR_{\tau\nu}$
is determined by the deviation from maximal mixing for $s_{23}^2$.
 However,
one has the result that
$s_{23}^2>0.5$ leads to $\BR_{\tau\nu} > \BR_{\mu\nu}$
while $s_{23}^2<0.5$ leads to $\BR_{\tau\nu} < \BR_{\mu\nu}$,
which are opposite behaviours to those for the normal mass ordering.
This result was not explicitly pointed out
in Refs.~\cite{Perez:2008ha,Davidson:2009ha},
and is due to the fact that
dominant contributions of $s_{23}^2$
to $(m_L^\dagger m_L^{})_{\mu\mu}$ and $(m_L^\dagger m_L^{})_{\tau\tau}$
come with $\Delta m^2_{31}$,
whose sign is flipped depending on the neutrino mass ordering.

%%%%%%  ratio of BRs  %%%%%%%%
 We now study the numerical value of the ratio of $\BR_{e\nu}$ and $\BR_{\mu\nu}$
as a function of $m_0$, 
for various values of $\sin^2{2\theta_{13}}$ and $s_{23}^2$.
The ratio does not change even if other decay channels
(such as $H^\pm\to W^\pm Z$ for $v_\Delta~(v_\nu) \gsim 0.1\,\MeV$
and $H^\pm\to W^\pm H^0$ for $m_{H^\pm}^{} > m_{H^0}^{}$)
have significant BRs.
 We note that the cross section for
$q\overline q \to H^+H^-$ depends on $m_{H^\pm}$, 
and approximate information on $m_{H^\pm}$
can be obtained from the $M_{T2}$ distribution of the signal,
as shown in Ref.~\cite{Davidson:2010sf}.
 However, given the sizeable uncertainty
in the extraction of $m_{H^\pm}$
we propose to use the ratio of $\BR_{e\nu}$ and $\BR_{\mu\nu}$
in which this uncertainty essentially cancels out,
thus enabling a more precise determination of the neutrino parameters.
 We note that there was no (explicit) quantitative study 
of this ratio in Refs.%
~\cite{Davidson:2009ha,Davidson:2010sf,Perez:2008ha},
although a qualitative discussion was given
in Ref.~\cite{Davidson:2010sf}.
 In Fig.~\ref{Fig:BRrate}~(upper panel) 
we show $\BR_{e\nu}/\BR_{\mu\nu}$ for a normal mass ordering.
 The central grey region corresponds to $s^2_{23}=0.5$ and
$0.07 < \sin^2{2\theta_{13}} < 0.11$.
 The dashed line (dot-dashed line) corresponds to
the largest (smallest) value of $\BR_{e\nu}/\BR_{\mu\nu}$ for a given $m_0$,
and is obtained for $s^2_{23}=0.4(0.6)$ and
$\sin^2{2\theta_{13}}=0.11 (0.07)$.
 As expected,
one can see that $\sin^2{2\theta_{13}}$
causes the most uncertainty
in $\BR_{e\nu}/\BR_{\mu\nu}$ for smaller values of $m_0$,
while $s^2_{23}$ gives the most uncertainty for larger $m_0$.
Since the ratio changes monotonically in a wide range
($0.05 \lesssim \BR_{e\nu}/\BR_{\mu\nu} \lesssim 0.9$)
with respect to $m_0$, a measurement
of $\BR_{e\nu}/\BR_{\mu\nu}$ would determine the value of $m_0$,
which might be more difficult to obtain from $H^{\pm\pm}$ decays alone
due to the additional uncertainty from the Majorana phases. For example,
$\BR_{e\nu}/\BR_{\mu\nu} \simeq 0.3$
means $m_0 \simeq 0.02\,\eV$ and $\Delta m^2_{31} > 0$.
 In Fig.~\ref{Fig:BRrate}~(lower panel)
we show $\BR_{\mu\nu}/\BR_{e\nu}$
(i.e.\ the inverse of  the ratio plotted
in the upper panel of Fig.~\ref{Fig:BRrate})
for an inverted mass ordering.
 Again,
the central grey region corresponds to $s^2_{23}=0.5$ and
$0.07 < \sin^2{2\theta_{13}}< 0.11$.
 As expected,
varying $\sin^2{2\theta_{13}}$ has very little effect
on the ratio {$\BR_{\mu\nu}/\BR_{e\nu}$
for an inverted mass ordering.
 The maximum (minimum) value of
  $\BR_{\mu\nu}/\BR_{e\nu}$ with a fixed $m_0$ is again obtained for $s^2_{23}=0.4 (0.6)$ and 
$\sin^2{2\theta_{13}}=0.11 (0.07)$.
 Hence a precise measurement of this
ratio would provide simultaneous information on
$s^2_{23}$, $m_0$ and the neutrino mass ordering.
For example,
$\BR_{\mu\nu}/\BR_{e\nu} < 0.5$ indicates
$s_{23}^2 > 0.5$, $m_0 \lesssim 0.01\,\eV$, and $\Delta m^2_{31} < 0$.
 If  the ratio in a range $0.65\,\text{-}\,0.9$ is observed
one can obtain a lower bound on $m_0~(= m_3)$.

%--------------------------------
\begin{figure}[t]
\begin{center}
\includegraphics[origin=c, angle=-90, scale=0.4]{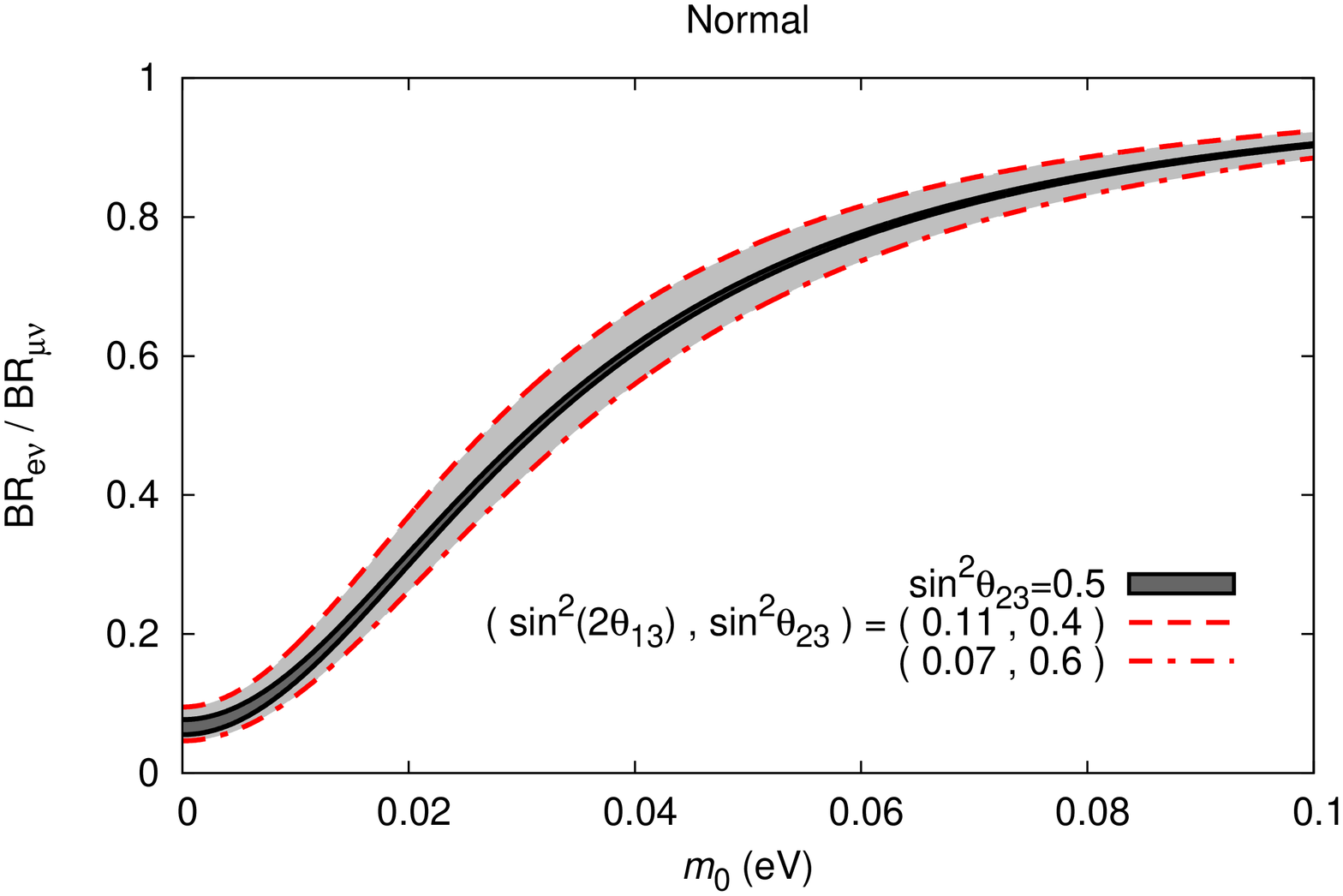}
\\\vspace*{-15mm}
\includegraphics[origin=c, angle=-90, scale=0.4]{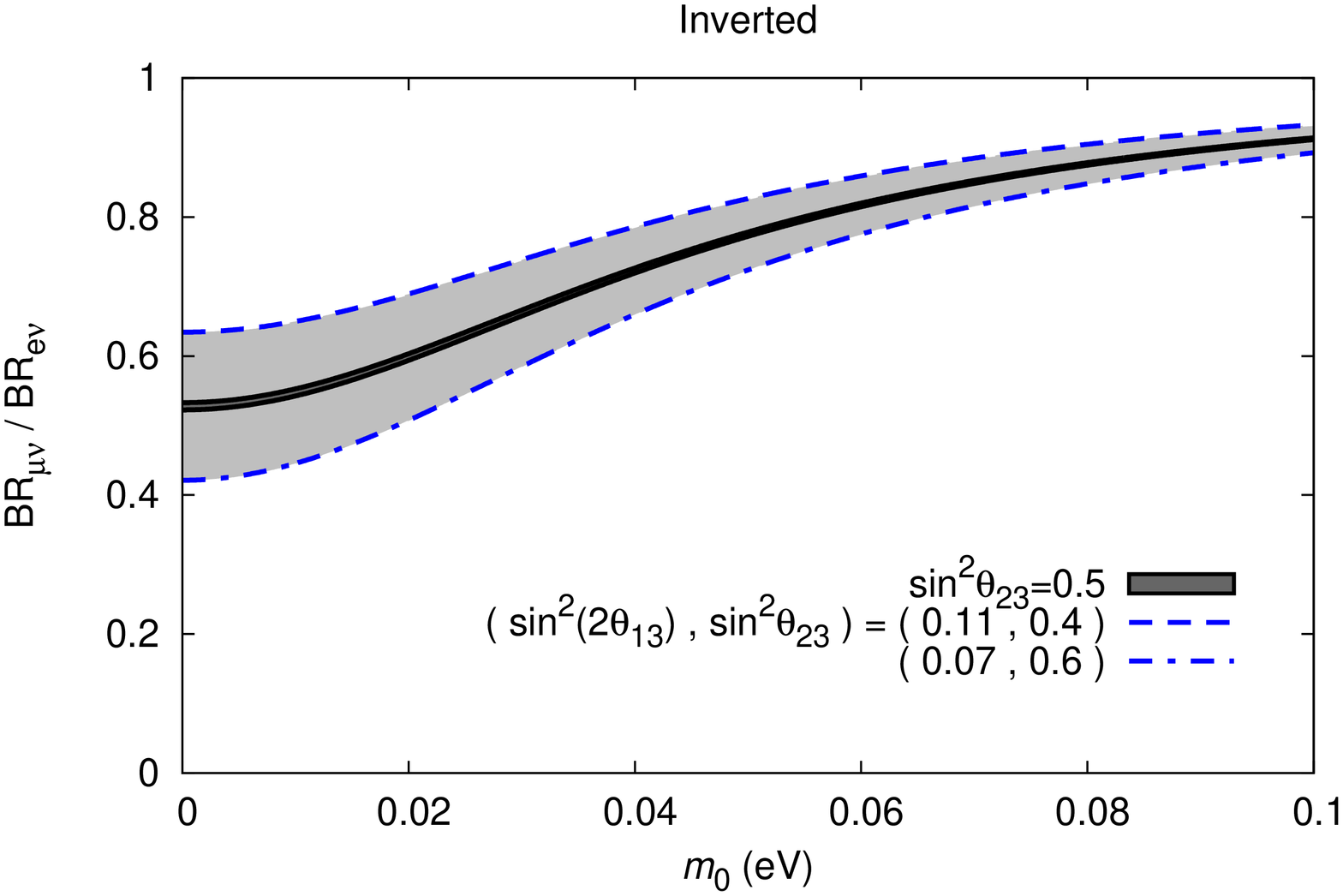}
\vspace*{-20mm}
\caption{
 Upper panel: the ratio $\BR_{e\nu}/\BR_{\mu\nu}$
as a function of $m_0(=m_1)$, 
for various values of $\sin^2{2\theta_{13}}$
and $s_{23}^2$ and a normal mass ordering.
 Lower panel: $\BR_{\mu\nu}/\BR_{e\nu}$
as a function of $m_0(=m_3)$ for an inverted mass ordering.}
\label{Fig:BRrate}
\end{center}
\end{figure}
%--------------------------------

We now discuss the phenomenology of $H^\pm$ at the LHC
by applying the above results to the phenomenological discussion
already given in Ref.~\cite{Davidson:2010sf}.
 In the $\nu$2HDM
the main production process of $H^\pm$ is
via $q\overline q\to \gamma, Z\to H^+H^-$.
 A simulation of the detection prospects of this process
has been performed in Ref.~\cite{Davidson:2010sf},
in which the signatures
$H^+H^-\to e^+e^-\nu\nu$, $e^\pm\mu^\mp\nu\nu$ and $\mu^\pm\mu^\mp\nu\nu$
were studied.
Detection prospects are best for the case of
an inverted neutrino mass ordering,
because the sum of $\BR_{e\nu}$ and $\BR_{\mu\nu}$
is always above $60\%$,
while for the case of a normal mass ordering
this sum of BRs can drop as low as $40\%$.
 By combining results for all three channels
($e^+e^-\nu\nu$, $e^\pm\mu^\mp\nu\nu$ and $\mu^+ \mu^- \nu\nu$),
detection at the $5\sigma$ level
in the $\nu$2HDM
for any choice of mass spectrum and mixing parameters
was shown to be possible for $m_{H^\pm}=100\,\GeV$~($300\,\GeV$)
with between $20\,\fb^{-1}$ and $80\,\fb^{-1}$
($57\,\fb^{-1}$ and $450\,\fb^{-1}$) of integrated luminosity
at $\sqrt s=14\,\TeV$~\cite{Davidson:2010sf}. Thus a signal could be possible
in the early stages of the $\sqrt s=14\,\TeV$ run of the LHC\@.
 In the HTM,
$2.7$ times larger integrated luminosity is required
because of the different $I_3$.

For the region of $m_0<10^{-3}\,\eV$
where the exact value of  $\sin^2{2\theta_{13}}$
plays an important role for a normal mass ordering,
the first signal of $H^+H^-$ with $m_{H^\pm}^{}=100\,\GeV$
would come in the channel $\mu^+ \mu^- \nu\nu$
(for which between $10\,\fb^{-1}$ and $80\,\fb^{-1}$
of integrated luminosity would be necessary
in the $\nu$2HDM).
 The small value of $\BR_{e\nu}$ for $m_0<10^{-3}\,\eV$ ensures that
detection of the $H^+H^-\to e^+e^-\nu\nu$ signal
would require very large ($> 10^4\,\fb^{-1}$) integrated luminosities,
which are possibly beyond the reach of an upgraded LHC\@.
 Reference~\cite{Davidson:2010sf} states that
the detection of the channel $e^\pm\mu^\mp\nu\nu$
for $m_{H^\pm}^{}=100\,\GeV$
would require integrated luminosities $\simeq 650\,\fb^{-1}$
in an optimistic case of $\sin^2{2\theta_{13}} \simeq 0.12$
(a region $0 \leq \sin^2{2\theta_{13}} \lesssim 0.12$
is used in Ref.~\cite{Davidson:2010sf}).
 In a pessimistic case $\sin^2{2\theta_{13}}= 0$,
integrated luminosities of a $\text{several}\times 10^3\,\fb^{-1}$
were required in the $\nu$2HDM
%maybe 2500 fb^-1 = 650 fb^-1 x (3.8/1)
because of a smaller $\BR_{e\nu}$.
 However, as already shown in Fig.~\ref{Fig:BRs-N},
the lower bound on $\BR_{e\nu}$
have now been improved by a factor of three
by virtue of the recent measurement of $\sin^2{2\theta_{13}}$.
 The required luminosity
to obtain a signal for $e^\pm\mu^\mp\nu\nu$ for $m_0<10^{-3}\,\eV$
has now been reduced to about $1000\,\fb^{-1}$
%maybe 900fb^-1 = 650 fb^-1 x (3.8/2.7)
even in a pessimistic case
in the $\nu$2HDM,
which is well within the reach of an upgraded LHC\@. Of course, for $m_0>10^{-3}\,\eV$
one sees from Fig.~\ref{Fig:BRs-N} that $\BR_{e\nu}$ starts to increase up to its maximum
value of BR$\sim 30\%$, and thus signals in all three channels
($H^+H^-\to e^+e^-\nu\nu$, $e^\pm\mu^\mp\nu\nu$ and $\mu^\pm\mu^\mp\nu\nu$) would become a possibility
with the envisaged integrated luminosities of the LHC\@. 

The exact value of $s_{23}^2$ plays a crucial role in determining
how much integrated luminosity is required for discovery of $H^\pm$,
because this parameter has a large effect on $\BR_{\mu\nu}$
(which is easier to detect) and $\BR_{\tau\nu}$,
unless the neutrinos are quasi-degenerate.
 If $\sin^2{2\theta_{23}} \simeq 1$ is precisely verified
by long baseline experiments in the near future,
then such a scenario would act to improve  the
predictions of $\BR_{\mu\nu}$ and $\BR_{\tau\nu}$.
 Alternatively,
if a significant deviation from $\sin^2{2\theta_{23}} = 1$
has been measured by long baseline experiments
and $H^\pm$ of the HTM or $\nu$2HDM has been discovered at the LHC,
then a measurement of $\BR(H^\pm\to \mu^\pm\nu)/\BR(H^\pm\to e^\pm\nu)$
could provide information on $s^2_{23}$
(and the sign of $\Delta m^2_{31}$)
earlier than oscillation experiments,
thereby removing the octant degeneracy.
Such information would be helpful
for CP-violation searches in future oscillation experiments.

\section{Conclusions}

 We have studied the branching ratios~(BRs)
of $H^\pm\to e^\pm\nu$, $H^\pm\to \mu^\pm\nu$ and $H^\pm\to \tau^\pm\nu$
in the context of the Higgs Triplet Model
and the neutrinophilic Two-Higgs-Doublet Model.
 We went beyond the analyses of previous papers
by quantifying the individual effect of the neutrino mixing angles
$\theta_{13}$ and $\theta_{23}$ on the above BRs.
We showed that the recent measurement of
$\sin^2{2\theta_{13}}=0.07\,\text{-}\,0.11$
has important implications for $\BR(H^\pm\to e^\pm\nu)$
in the case of a normal neutrino mass ordering.
 The above measurement of $\sin^2{2\theta_{13}}$
rules out (at about 95\%~c.l.) the previously allowed region of
$1\% < \BR(H^\pm\to e^\pm\nu) < 2.7\%$,
while constraining the BR to lie in the region
$2.7\% < \BR(H^\pm\to e^\pm\nu) < 30\%$.
 This ensures that integrated luminosities of
about $1000\,\fb^{-1}$~($2700\,\fb^{-1}$)
should be enough to observe a signal for
$q\overline q\to H^+H^-\to e^\pm \mu^\mp\nu\nu$
in the $\nu$2HDM~(HTM)
at the upgraded LHC even if $m_0< 10^{-3}\,\eV$, where $\BR(H^\pm\to e^\pm\nu)$ has a minimum value.

We also showed that
$\BR(H^\pm \to \mu^\pm\nu)$ and $\BR(H^\pm \to \tau^\pm\nu)$
can deviate by up to $20\%$ depending on the value of $s_{23}^2$.
 For the case of $s_{23}^2>0.5$ and a normal mass ordering
one has the result
$\BR(H^\pm \to \mu^\pm\nu) > \BR(H^\pm \to \tau^\pm\nu)$,
while for $s_{23}^2<0.5$ one has
$\BR(H^\pm \to \mu^\pm\nu) < \BR(H^\pm \to \tau^\pm\nu)$.
For the case of an inverted neutrino mass ordering
one has the converse results.
 We proposed to use the ratio of $\BR_{e\nu}$ and $\BR_{\mu\nu}$
in which the uncertainty from $m_{H^\pm}$
in the production cross section cancels out,
thus enabling a more precise determination of the neutrino parameters
than for the cases of using $\BR_{e\nu}$ and $\BR_{\mu\nu}$ alone.
 Accurate information on $s_{23}$, $m_0$ and the neutrino mass ordering
could then be obtained,
some of which might be difficult ($m_0$ is impossible)
to obtain at future neutrino oscillation experiments.
 Such information would be helpful
for CP-violation searches in future oscillation experiments.

\section*{Acknowledgements}
 We thank Koji Tsumura for a useful comment.
 S.M.\ is partially supported through the NExT institute. The work of H.S.\ was supported in part
by JSPS KAKENHI Grant Number~23740210.

\end{document}